\documentclass[3p,times,procedia]{elsarticle}
\usepackage{nupha_ecrc}

\volume{00}

\firstpage{1}

\journalname{Nuclear Physics A}

\runauth{}

\jid{nupha}

\jnltitlelogo{Nuclear Physics A}

\usepackage{amssymb}

\usepackage{lineno}

\usepackage[figuresright]{rotating}

\begin{document}

\begin{frontmatter}



\dochead{XXVIIIth International Conference on Ultrarelativistic Nucleus-Nucleus Collisions\\ (Quark Matter 2019)}

\title{Upgrading the Inner Tracking System and the Time Projection Chamber of ALICE}


\author{Felix Reidt\\ on behalf of the ALICE collaboration}

\address{CERN, 1211 Geneva 23, Switzerland}

\begin{abstract}
The ALICE experiment at CERN is undergoing a major upgrade during the Long Shutdown 2 (LS2) of the LHC during 2019-2020. The key elements regarding the central barrel are the installation of a new Inner Tracking System (ITS) and the upgrade of the large Time Projection Chamber (TPC). The TPC is currently being upgraded with a new readout system, including new GEM-based Readout Chambers and new front-end electronics enabling us to operate it in continuous mode. The new ITS based on CMOS Monolithic Active Pixel Sensors will significantly improve the impact parameter resolution and the tracking efficiency, especially for particles with low transverse momenta, as well as the readout-rate capability. In the following, the upgrades and their assembly and commissioning status will be outlined.

Furthermore, an outlook will be given on plans for the upgrade of the ITS foreseen for the third long shutdown of the LHC in \mbox{2025-2026}. For this upgrade, the three innermost layers could be replaced by cylindrical detector layers, made of curved wafer-scale-sized CMOS sensors, to lower even further the material budget and to improve the pointing resolution.
\end{abstract}


\end{frontmatter}


\section{Introduction}
\label{}
The ALICE experiment~\cite{Aamodt2008} is upgrading its experimental apparatus during the currently ongoing Long Shutdown 2 (LS2) of the LHC. ALICE will significantly improve its sensitivity to rare probes by fully recording Pb--Pb collision data at 50\,kHz interaction rate achievable after the accelerator upgrade, and increasing the minimum-bias data sample by about a factor of 100. Furthermore, the increased vertexing performance will allow for a significantly more effective identification and reconstruction of heavy-flavour hadron decays.

During LS2, the key upgrade elements for the central barrel are the installation of a new Inner Tracking System (ITS2)~\cite{Alice2014b} and the upgrade of the large Time Projection Chamber (TPC)~\cite{Alice2013b}. The ITS2 is equipped with Monolithic Active Pixel Sensors (MAPS). For the TPC, the Multi-Wire Proportional Chambers (MWPCs) readout were replaced by Gas Electron Multiplier (GEM) chambers. In the following, the current status of commissioning of the two detector systems will be outlined.

The next step will be the R\&D for the Inner Tracking System 3 (ITS3)~\cite{Alice2019}, planned for the Long Shutdown 3 (LS3) foreseen for 2025-2026. This novel detector will be based on 3 bent silicon layers reducing the material budget even further. The new detector concept will be briefly introduced in Sec.~\ref{sec:its3}.

\section{Time Projection Chamber Upgrade}
The TPC~\cite{Alice2013b} is the main tracking and Particle IDentification (PID) device of ALICE. Previously, the TPC was read out with MWPCs. A gating grid was used to prevent ion backflow, which limited its readout rate to a few kHz. In order to be able to record all Pb-Pb interactions at a rate of 50\,kHz, the MWPCs were replaced by GEMs. The new ReadOut Chambers (ROCs) will allow for a continuous readout and hence a full reconstruction of all collisions. 

Operation without a gating grid will lead to a substantial accumulation of space charge in the drift volume of the TPC, resulting in space charge distortions. In order to keep them to a minimum, a special configuration based on 4-GEM stacks with different hole pitches has been adopted: two large-pitch (280\,$\mu$m) GEM foils, embraced by two standard-pitch (140\,$\mu$m) GEM foils~\cite{Mathis:2018sjk,Gasik:2018mfu}. All ROCs including spares arrived at CERN in early 2019 and were tested for stability using X-ray irradiation or particles in the ALICE experimental cavern.

The TPC has been moved to surface for the installation of the new GEM ROCs in March 2019. During the course of 2019, the MWPC ROCs as well as the corresponding readout electronics and services have been deinstalled and the GEM ROCs along with the corresponding readout electronics have been installed. In December 2019, the pre-commissioning on surface with cosmic muons, laser, pulser and X-ray irradiation will start. This will allow us to validate the ROCs before the detector will be moved in the experimental cavern in March 2020.

\begin{figure}
    \centering
    \includegraphics[height=3.5cm]{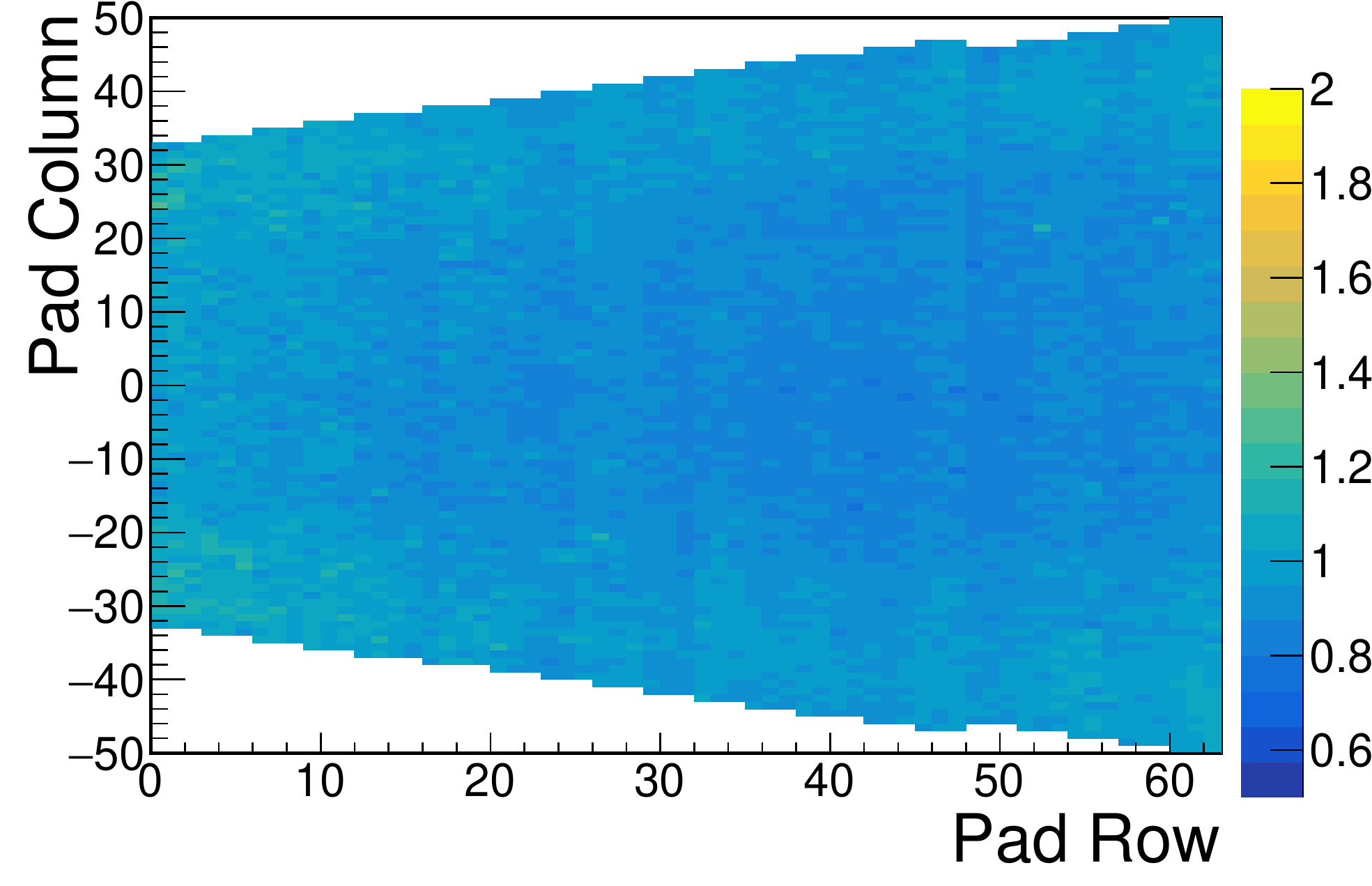}
    \hspace{0.1cm}
    \includegraphics[height=3.5cm]{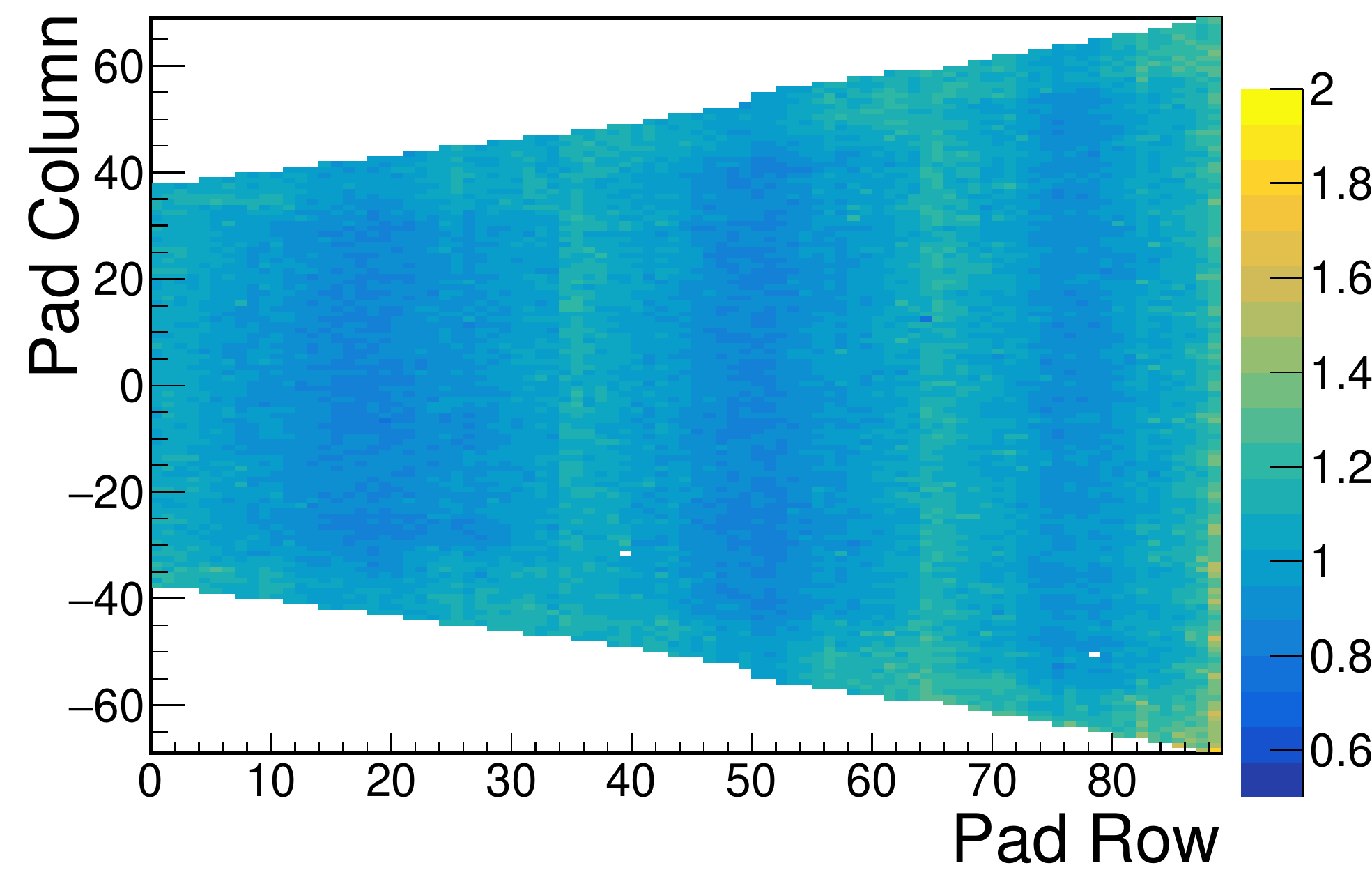}
    \hspace{0.1cm}
    \includegraphics[height=3.5cm]{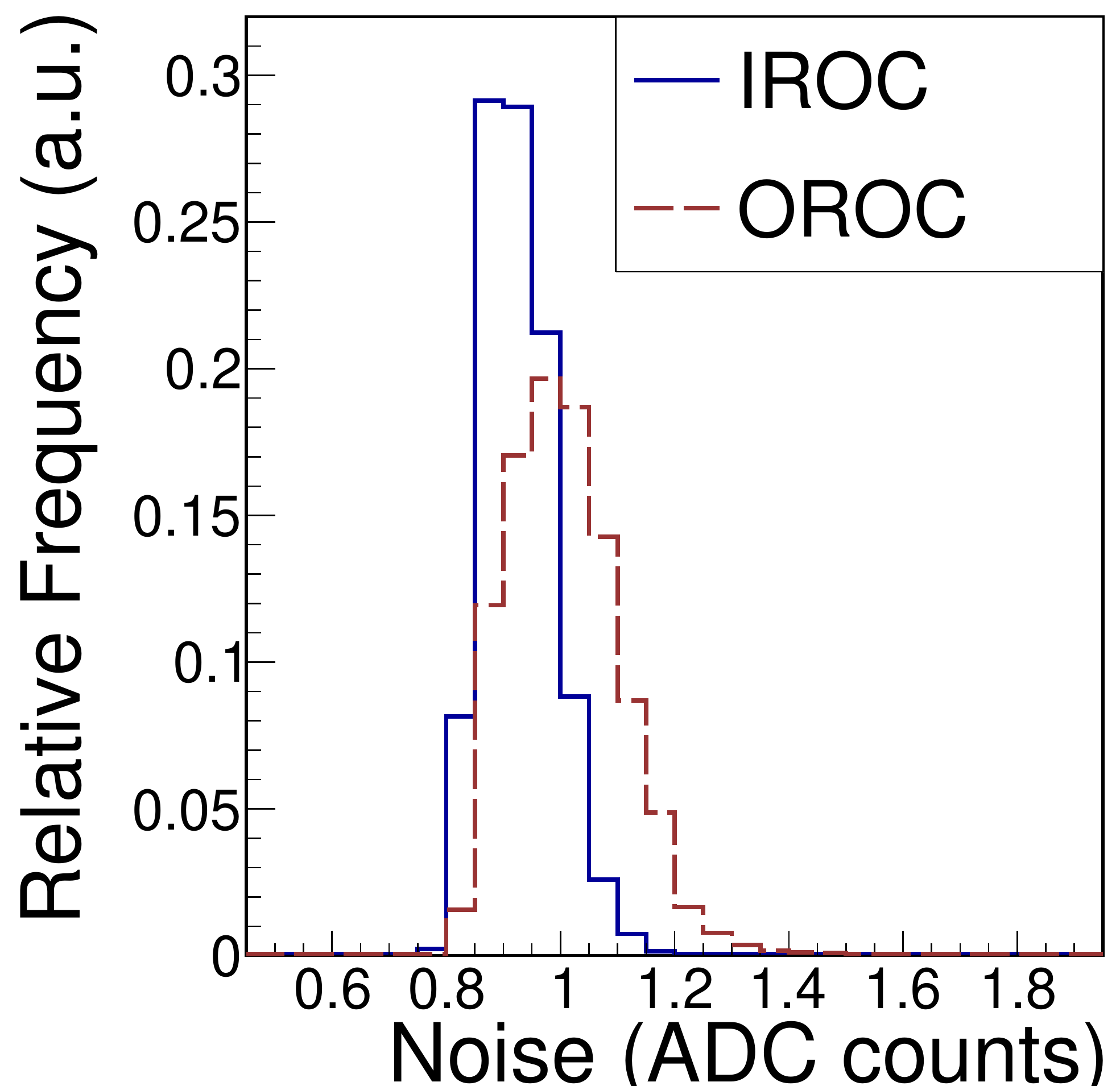}
    \caption{Noise performance of TPC readout chambers on one TPC sector. From left to right: IROC noise map, OROC noise map, IROC noise distribution (blue, solid line) and OROC noise distribution (red, dashed line).}
    \label{fig:TPCnoisePerf}
\end{figure}

In Fig.~\ref{fig:TPCnoisePerf}, the noise maps of an Inner ROC (IROC, left) and Outer ROC (OROC, middle) and the corresponding noise distributions (right) are shown. The average noise on a given TPC sector is 0.93 and 1.01\,ADC~counts for the IROC and OROC, respectively. The tails are due to different path lengths from the readout pad to the front-end electronics leading to different capacitances. The overall performance is in line with the target value of 1\,ADC~count of noise corresponding to 670\,$e^{-}$. 

\section{Inner Tracking System Upgrade - ITS2}
The ITS2 will fully replace the previous ITS. It is based on Monolithic Active Pixel Sensors and has a total surface of 10\,m$^2$. It consists of 7 concentrically cylindrical layers ranging from radii of $2.3$\,cm to 40\,cm, with a material budget of $0.35\%$\,$X_0$ and $1\%$\,$X_0$ for the innermost 3 and outermost 4 layers, respectively. Layers 0 to 2 form together the Inner Barrel (IB) and layers 3 to 6 the Outer Barrel (OB). The OB is furthermore subdivided into Middle Layers (MLs), comprising layers 3 and 4, and Outer Layers (OLs), comprising layers 5 and 6.

The key building block is the ALPIDE Monolithic Active Pixel Sensor~\cite{AglieriRinella:2017lym,Martinengo:2017fuc}, which features in-pixel amplification, discrimination and multiple hit buffers as well as in-matrix data sparsification. For the IB, modules consisting of 9 chips, thinned down to 50\,$\mu$m, form a stave. For the OB, modules of 14 chips with a thickness of 100\,$\mu$m each are assembled into staves. A ML stave consist of two half-staves of 4 modules each. An OL stave, consist of a total of 14 modules distributed over 2 half-staves. The full ITS2 consists of 48 IB, 54 ML and 90 OL staves.

From the end of 2016 until the end of 2019, a total of 72000 ALPIDE sensors have been produced and tested. The overall chip production yield amounts to $63.9\%$. For the IB modules, only the best category of chips has been used, which has a yield of $30.4\%$. A total of about 3000 modules have been produced reaching a yield of $71.5\%$ and  $84.2\%$ for IB and OB modules, respectively. During the second half of 2018 until fall 2019, all seven detector layers have been assembled and installed in the half-barrels in an assembly clean room at CERN . 

\begin{figure}
    \centering
    \includegraphics[height=5.5cm]{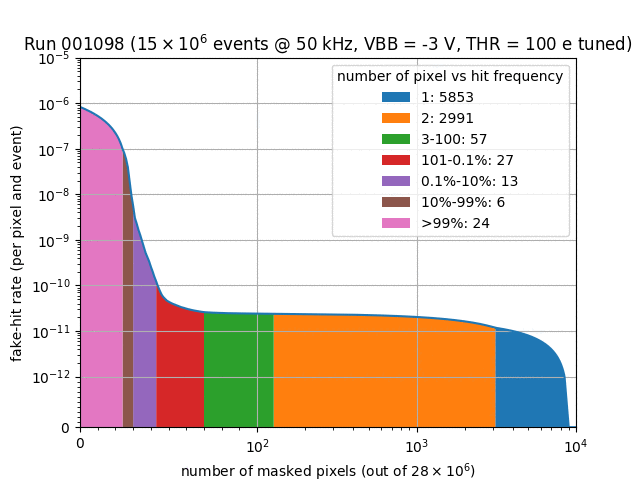}
    \includegraphics[height=5.5cm]{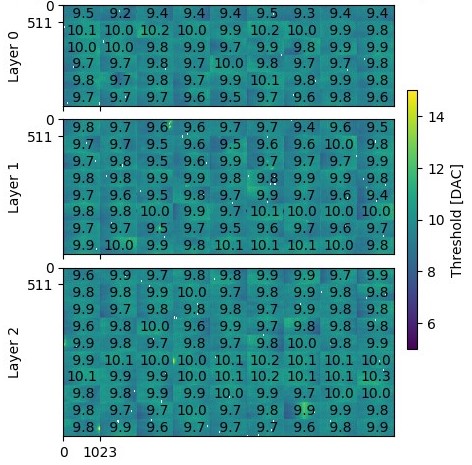}
    \caption{Results from the ITS2 IB commissioning: fake-hit rate of 54 chips (left), threshold map of an Inner Barrel Half-Barrel after tuning (right).}
    \label{fig:ITS2commissioningResults}
\end{figure}

Furthermore, all detector services including cooling plant, power distribution, readout electronics and computing system have been installed in and around the assembly clean room to allow for full commissioning of the detector on surface. As first results, fake-hit rates and threshold measurements have been carried out. In Fig.~\ref{fig:ITS2commissioningResults}, left, the fake-hit rate for 54 chips in half-layer 0 is shown to be below $10^{-10}$ fake hits per pixel and event when masking only about 50 out of $28 \times 10^{6}$ pixels. This is significantly better than the required $10^{-6}$ fake hits per pixel and event. In the right half of Fig.~\ref{fig:ITS2commissioningResults}, a threshold map for an IB Half-Barrel after tuning the threshold to 10\,DAC counts corresponding to 100\,$e^{-}$ is shown. The overall uniformity is excellent and the tuning procedure has been validated. Furthermore, first cosmic-ray tracks have been observed in the detector.

\section{Inner Tracking System Upgrade - ITS3}\label{sec:its3}
\begin{figure}
    \centering
    \includegraphics[height=5cm]{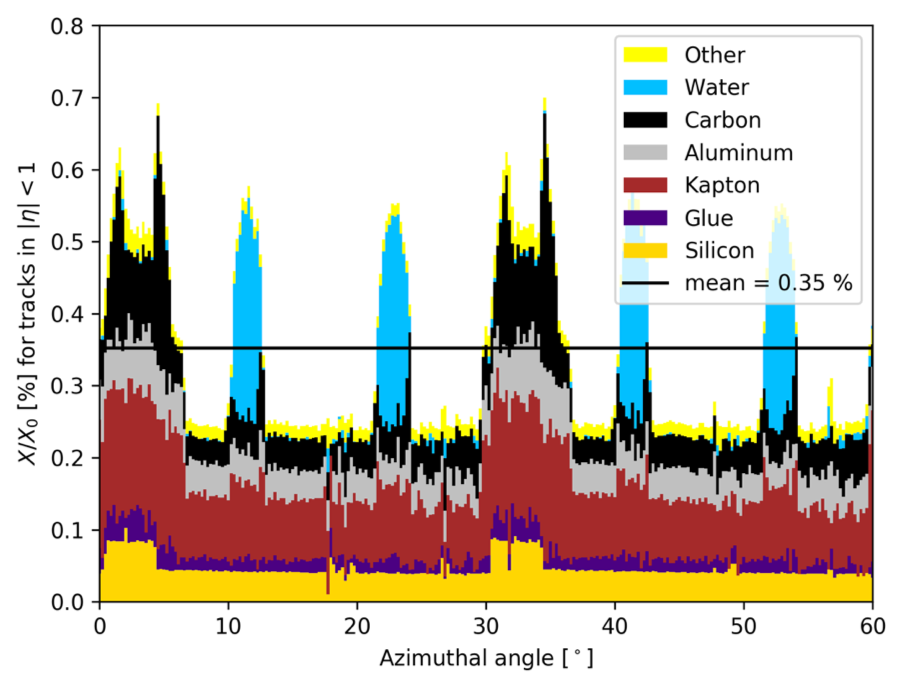}
    \hspace{0.5cm}
    \includegraphics[height=5cm]{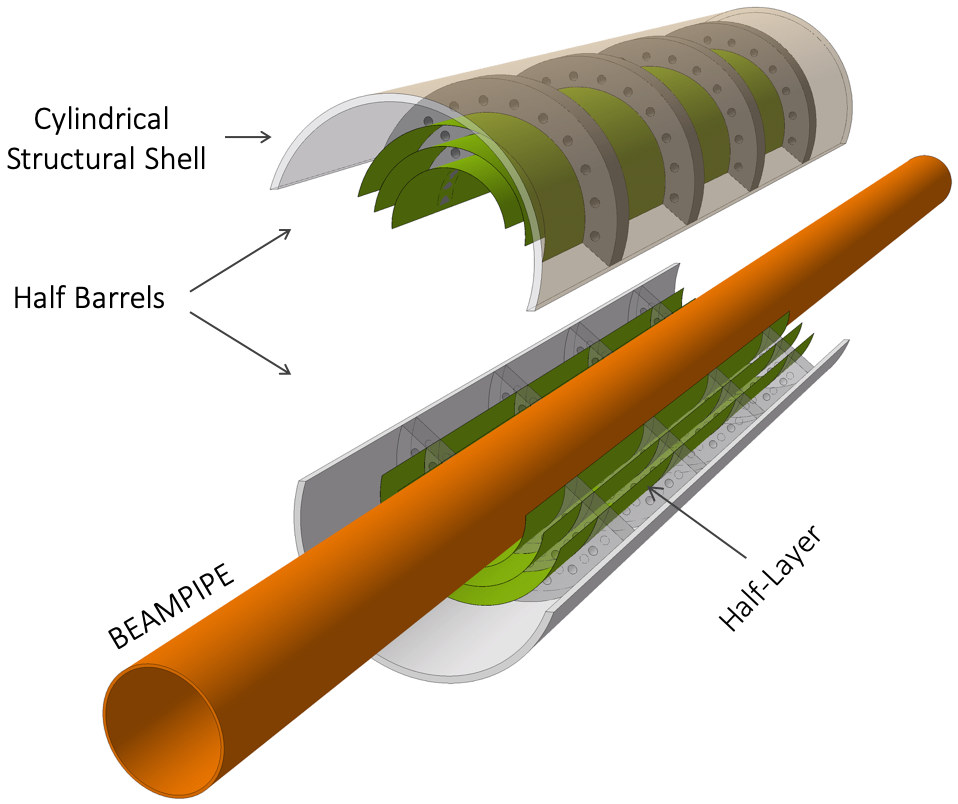}
    \caption{Material budget of an ITS2 IB layer (left), schematic drawing of ITS3 (right)}
    \label{fig:ITS3}
\end{figure}

The ITS2 material budget is largely dominated by support and electrical substrate (cf.\ Fig.~\ref{fig:ITS3}, left) and only about $15\%$ is owing to the silicon itself. Furthermore, the material distribution is very irregular due to the overlap of staves, the cooling pipes, as well as the support structures. The novel concept of the ITS3~\cite{Alice2019} is the introduction of 3 layers of curved silicon sensors (cf.\ Fig.~\ref{fig:ITS3}, right), which should replace the 3 ITS2 IB layers. A reduction of the power consumption below 20\,$\textrm{mW/cm}^{2}$ for the active area will allow to rely on air cooling. The mechanical support can be removed by exploiting the increased stiffness of the rolled silicon wafers. The electrical substrate will be integrated directly into the chip, which would span a major fraction of a layer.

The main building block of the ITS3 is an ultra-thin (20\,$\mu$m to 40\,$\mu$m), wafer-scale (300\,mm) CMOS Monolithic Active Pixel Sensor whose dimensions reach up to 280\,mm by 94\,mm. Owing to the flexibility of silicon at such thicknesses, wafer-size sensors will be bent into half-cylinders of radii of 18, 24, and 30\,mm, respectively, to form the new concentric layers. The power intensive periphery of such scale sensors will placed outside the fiducial volume, where traditional cooling can be implemented.

The new detector will go together with a beam pipe with smaller diameter as well as reduced wall thickness. The total material up to a radius of 4\,cm will reduce by about a factor of 3 and it will be more homogeneously distributed. This should allow for a reduction of the systematic uncertainty for those measurements that require a precise description of the material distribution.

The ITS3 is expected to improve the pointing resolution by a factor of 2 compared to ITS2 and further improve the tracking efficiency at low momenta with a subsequent improvement of the physics performance for heavy-flavour baryon and low-mass dielectron measurements.

\section{Summary and Outlook}
The preparation of the upgrades of TPC and ITS during LS2 is on track for operation in 2021. The TPC ROCs and electronics perform as expected. Results from the ITS2 on-surface commissioning are confirming the expected performance. Both TPC and ITS2 are set to run at 50\,kHz in Run 3.\\
The proposed ITS3 detector will have unprecedented performance approaching the limit of a massless detector. A major part of the ALICE physics program will benefit from it and new channels are expected to become accessible.




\bibliographystyle{elsarticle-num}
\bibliography{literature.bib}

\end{document}